\pdfoutput=1
\documentclass[12pt,a4paper]{article}
\usepackage{ifthen} 
\newboolean{pdflatex}
\setboolean{pdflatex}{true} 

\newboolean{articletitles}
\setboolean{articletitles}{true} 

\newboolean{uprightparticles}
\setboolean{uprightparticles}{false} 

\newboolean{inbibliography}
\setboolean{inbibliography}{false} 

\def\BdKsMuMu     {\decay{\Bd}{ \KS \mup\mun}}
\def\BdKzMuMu     {\decay{\Bd}{ \Kz \mup\mun}}
\def\BdJpsiKs     {\decay{\Bd}{ \jpsi \KS}}
\def\BdJpsiKz     {\decay{\Bd}{ \jpsi K^{0}}}
\def\BuJpsiK       {\decay{\Bp}{ \jpsi \Kp}}
\def\BuKMuMu     {\decay{\Bp}{ \Kp \mup\mun}}

\def\BuKstMuMu     {\decay{\Bp}{ K^{*+} \mup\mun}}
\def\BuJpsiKst    {\decay{\Bp}{ \jpsi K^{*+}}}
\def\BdKstMuMu     {\decay{\Bd}{ \Kstarz \mup\mun}}

\def\BdJpsiKst    {\decay{\Bd}{ \jpsi K^{*0} }}
\def\BKMuMu    {\decay{\B}{ K \mup\mun}}
\def\BKstMuMu {\decay{\B}{ K^{*} \mup \mun}}
\def\BdKstarzMuMu {\decay{\Bd}{ K^{*0} \mup \mun}}
\def\BuKstarpMuMu {\decay{\Bu}{ K^{*+} \mup \mun}}

\def\BdJpsimumuKstarz  {\decay{\Bd}{ (\jpsi\to\mup\mun) \Kstarz}}
\def\AllModes    {\decay{\B}{ K^{(*)} \mup\mun}}
\def\CharModes    {\decay{\Bu}{ K^{(*)+} \mup\mun}}
\def\NeutModes    {\decay{\Bd}{ K^{(*)0} \mup\mun}}

\def\qq {\ensuremath{q^{2}}\xspace}
\def\AI {\ensuremath{A_{\rm I}}\xspace}

\def\JpsiModes {\decay{\B}{ \jpsi K^{(*)}  }}

\usepackage{microtype}
\usepackage{lineno}  
\usepackage{xspace} 

\usepackage{subfigure}
\usepackage{graphicx}  
\usepackage{color}
\usepackage{colortbl}
\graphicspath{{./figs/}} 

\usepackage{amsmath} 
\usepackage{amssymb}
\usepackage{amsfonts}
\usepackage{upgreek} 

\newcommand*\patchAmsMathEnvironmentForLineno[1]{%
\expandafter\let\csname old#1\expandafter\endcsname\csname #1\endcsname
\expandafter\let\csname oldend#1\expandafter\endcsname\csname
end#1\endcsname
 \renewenvironment{#1}%
   {\linenomath\csname old#1\endcsname}%
   {\csname oldend#1\endcsname\endlinenomath}%
}
\newcommand*\patchBothAmsMathEnvironmentsForLineno[1]{%
  \patchAmsMathEnvironmentForLineno{#1}%
  \patchAmsMathEnvironmentForLineno{#1*}%
}
\AtBeginDocument{%
\patchBothAmsMathEnvironmentsForLineno{equation}%
\patchBothAmsMathEnvironmentsForLineno{align}%
\patchBothAmsMathEnvironmentsForLineno{flalign}%
\patchBothAmsMathEnvironmentsForLineno{alignat}%
\patchBothAmsMathEnvironmentsForLineno{gather}%
\patchBothAmsMathEnvironmentsForLineno{multline}%
}

\usepackage{hyperref}    
\usepackage[all]{hypcap} 
\usepackage{ctable}

\def\lhcb {\mbox{LHCb}\xspace}

\ifthenelse{\boolean{uprightparticles}}%
{

 \def\Pmu         {\ensuremath{\upmu}\xspace}

 \def\Ppi         {\ensuremath{\uppi}\xspace}

 \def\Ppsi        {\ensuremath{\uppsi}\xspace}

 \def\PDelta      {\ensuremath{\Delta}\xspace}                 
 \def\PXi      {\ensuremath{\Xi}\xspace}                 
 \def\PLambda      {\ensuremath{\Lambda}\xspace}                 
 \def\PSigma      {\ensuremath{\Sigma}\xspace}                 
 \def\POmega      {\ensuremath{\Omega}\xspace}                 
 \def\PUpsilon      {\ensuremath{\Upsilon}\xspace}                 
 
 \def\PB      {\ensuremath{\mathrm{B}}\xspace}                 
                  
 \def\PD      {\ensuremath{\mathrm{D}}\xspace}

 \def\PJ      {\ensuremath{\mathrm{J}}\xspace}                 
 \def\PK      {\ensuremath{\mathrm{K}}\xspace}

 \def\Pb      {\ensuremath{\mathrm{b}}\xspace}                 
 \def\Pc      {\ensuremath{\mathrm{c}}\xspace}

 \def\Pi      {\ensuremath{\mathrm{i}}\xspace}

 \def\Ps      {\ensuremath{\mathrm{s}}\xspace}

}
{

 \def\Pmu         {\ensuremath{\mu}\xspace}

 \def\Ppi         {\ensuremath{\pi}\xspace}

 \def\Ppsi        {\ensuremath{\psi}\xspace}                 
                  
 \mathchardef\PDelta="7101
 \mathchardef\PXi="7104
 \mathchardef\PLambda="7103
 \mathchardef\PSigma="7106
 \mathchardef\POmega="710A
 \mathchardef\PUpsilon="7107
                  
 \def\PB      {\ensuremath{B}\xspace}                 
                  
 \def\PD      {\ensuremath{D}\xspace}

 \def\PJ      {\ensuremath{J}\xspace}                 
 \def\PK      {\ensuremath{K}\xspace}

 \def\Pb      {\ensuremath{b}\xspace}                 
 \def\Pc      {\ensuremath{c}\xspace}

 \def\Pi      {\ensuremath{i}\xspace}

 \def\Ps      {\ensuremath{s}\xspace}

}

\def\mup        {\ensuremath{\Pmu^+}\xspace}
\def\mun        {\ensuremath{\Pmu^-}\xspace} 
\def\mumu       {\ensuremath{\Pmu^+\Pmu^-}\xspace}

\def\squark    {\ensuremath{\Ps}\xspace}

\def\cquark    {\ensuremath{\Pc}\xspace}

\def\bquark    {\ensuremath{\Pb}\xspace}

\def\pion  {\ensuremath{\Ppi}\xspace}
\def\piz   {\ensuremath{\pion^0}\xspace}

\def\pip   {\ensuremath{\pion^+}\xspace}
\def\pim   {\ensuremath{\pion^-}\xspace}

\def\kaon  {\ensuremath{\PK}\xspace}
  \def\Kbar  {\kern 0.2em\overline{\kern -0.2em \PK}{}\xspace}

\def\Kz    {\ensuremath{\kaon^0}\xspace}

\def\Kp    {\ensuremath{\kaon^+}\xspace}
\def\Km    {\ensuremath{\kaon^-}\xspace}

\def\KS    {\ensuremath{\kaon^0_{\rm\scriptscriptstyle S}}\xspace} 
\def\KL    {\ensuremath{\kaon^0_{\rm\scriptscriptstyle L}}\xspace} 
\def\Kstarz  {\ensuremath{\kaon^{*0}}\xspace}
\def\Kstarzb {\ensuremath{\Kbar^{*0}}\xspace}
\def\Kstar   {\ensuremath{\kaon^*}\xspace}

\def\Kstarp  {\ensuremath{\kaon^{*+}}\xspace}

  \def\Dbar    {\kern 0.2em\overline{\kern -0.2em \PD}{}\xspace}
\def\D       {\ensuremath{\PD}\xspace}

\def\Dz      {\ensuremath{\D^0}\xspace}

\def\B       {\ensuremath{\PB}\xspace}
\def\Bbar    {\ensuremath{\kern 0.18em\overline{\kern -0.18em \PB}{}}\xspace}

\def\Bz      {\ensuremath{\B^0}\xspace}

\def\Bu      {\ensuremath{\B^+}\xspace}

\def\Bp      {\ensuremath{\Bu}\xspace}

\def\Bd      {\ensuremath{\B^0}\xspace}
\def\Bs      {\ensuremath{\B^0_\squark}\xspace}

\def\Bdb     {\ensuremath{\Bbar^0}\xspace}

\def\jpsi     {\ensuremath{{\PJ\mskip -3mu/\mskip -2mu\Ppsi\mskip 2mu}}\xspace}
\def\psitwos  {\ensuremath{\Ppsi{(2S)}}\xspace}

  \def\Y#1S{\ensuremath{\PUpsilon{(#1S)}}\xspace}

\def\Lz {\ensuremath{\PLambda}\xspace}
\def\Lbar {\ensuremath{\kern 0.1em\overline{\kern -0.1em\PLambda}}\xspace}

\def\Lb      {\ensuremath{\Lz^0_\bquark}\xspace}

\newcommand{\decay}[2]{\ensuremath{#1\!\to #2}\xspace}         

\def\to                 {\ensuremath{\rightarrow}\xspace}

\def\qsq       {\ensuremath{q^2}\xspace}

\def\CP                {\ensuremath{C\!P}\xspace}

\def\AT#1     {\ensuremath{A_{\mathrm{T}}^{#1}}\xspace}           

\def\C#1      {\ensuremath{\mathcal{C}_{#1}}\xspace}                       
\def\Cp#1     {\ensuremath{\mathcal{C}_{#1}^{'}}\xspace}                    
\def\Ceff#1   {\ensuremath{\mathcal{C}_{#1}^{\mathrm{(eff)}}}\xspace}        
\def\Cpeff#1  {\ensuremath{\mathcal{C}_{#1}^{'\mathrm{(eff)}}}\xspace}       
\def\Ope#1    {\ensuremath{\mathcal{O}_{#1}}\xspace}                       
\def\Opep#1   {\ensuremath{\mathcal{O}_{#1}^{'}}\xspace}                    

\newcommand{\tev}{\ifthenelse{\boolean{inbibliography}}{\ensuremath{~T\kern -0.05em eV}\xspace}{\ensuremath{\mathrm{\,Te\kern -0.1em V}}\xspace}}
\newcommand{\gev}{\ensuremath{\mathrm{\,Ge\kern -0.1em V}}\xspace}
\newcommand{\mev}{\ensuremath{\mathrm{\,Me\kern -0.1em V}}\xspace}
\newcommand{\kev}{\ensuremath{\mathrm{\,ke\kern -0.1em V}}\xspace}
\newcommand{\ev}{\ensuremath{\mathrm{\,e\kern -0.1em V}}\xspace}
\newcommand{\gevc}{\ensuremath{{\mathrm{\,Ge\kern -0.1em V\!/}c}}\xspace}
\newcommand{\mevc}{\ensuremath{{\mathrm{\,Me\kern -0.1em V\!/}c}}\xspace}
\newcommand{\gevcc}{\ensuremath{{\mathrm{\,Ge\kern -0.1em V\!/}c^2}}\xspace}
\newcommand{\gevgevcccc}{\ensuremath{{\mathrm{\,Ge\kern -0.1em V^2\!/}c^4}}\xspace}
\newcommand{\mevcc}{\ensuremath{{\mathrm{\,Me\kern -0.1em V\!/}c^2}}\xspace}

\def\mum  {\ensuremath{{\,\upmu\rm m}}\xspace}

\def\invfb   {\ensuremath{\mbox{\,fb}^{-1}}\xspace}

\def\ps   {\ensuremath{{\rm \,ps}}\xspace}

\newcommand{\chisq}{\ensuremath{\chi^2}\xspace}

\def\gsim{{~\raise.15em\hbox{$>$}\kern-.85em
          \lower.35em\hbox{$\sim$}~}\xspace}
\def\lsim{{~\raise.15em\hbox{$<$}\kern-.85em
          \lower.35em\hbox{$\sim$}~}\xspace}

\def\pt         {\mbox{$p_{\rm T}$}\xspace}

\def\mrad{\ensuremath{\rm \,mrad}\xspace}

\def\gauss      {\mbox{\textsc{Gauss}}\xspace}

\def\tell1  {TELL1\xspace}
\def\ukl1   {UKL1\xspace}

\usepackage{cite} 
\usepackage{mciteplus}

\begin{document}

\renewcommand{\thefootnote}{\fnsymbol{footnote}}
\setcounter{footnote}{1}


\begin{titlepage}
\pagenumbering{roman}

\vspace*{-1.5cm}
\centerline{\large EUROPEAN ORGANIZATION FOR NUCLEAR RESEARCH (CERN)}
\vspace*{1.5cm}
\hspace*{-0.5cm}
\begin{tabular*}{\linewidth}{lc@{\extracolsep{\fill}}r}
\ifthenelse{\boolean{pdflatex}}
{\vspace*{-2.7cm}\mbox{\!\!\!\includegraphics[width=.14\textwidth]{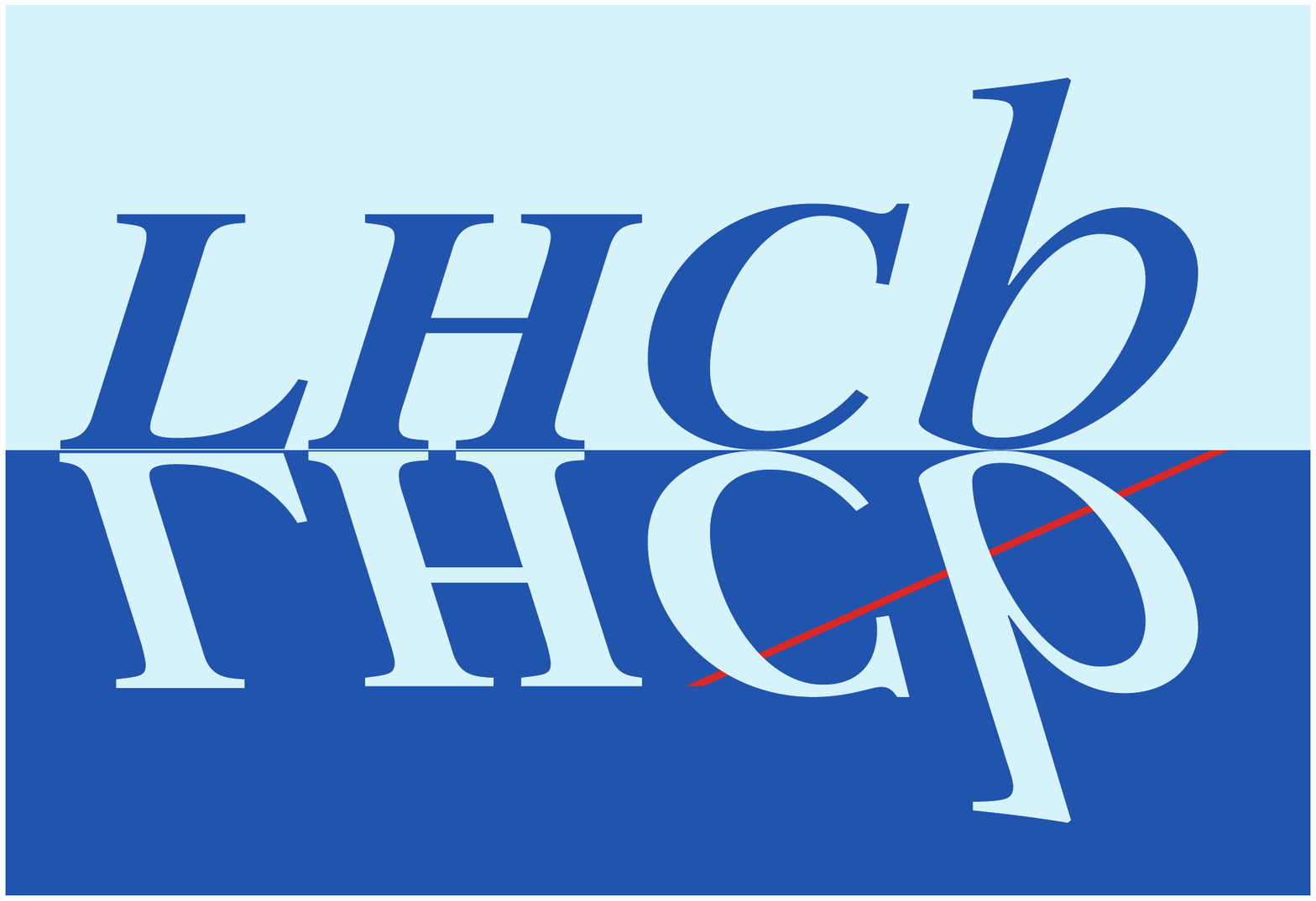}} & &}%
{\vspace*{-1.2cm}\mbox{\!\!\!\includegraphics[width=.12\textwidth]{lhcb-logo.eps}} & &}%
\\
 & & CERN-PH-EP-2014-055 \\  
 & & LHCb-PAPER-2014-006 \\  
 & & 23 Jun 2014 \\ 
\end{tabular*}

\vspace*{3.5cm}

{\bf\boldmath\huge
\begin{center}
Differential branching fractions \\ and isospin asymmetries \\ of \AllModes decays
\end{center}
}

\vspace*{2.0cm}

\begin{center}
The LHCb collaboration\footnote{Authors are listed on the following pages.}

\end{center}

\vspace{\fill}

\begin{abstract}
\noindent
The isospin asymmetries of $B \to K\mu^+\mu^-$ and $B \to
K^{*}\mu^+\mu^-$ decays and the partial branching fractions of the
$B^0 \to K^0\mu^+\mu^-$, $B^+ \to K^+\mu^+\mu^-$ and $B^+ \to
K^{*+}\mu^+\mu^-$ decays are measured as functions of the dimuon mass
squared, $q^2$. The data used correspond to an integrated luminosity
of 3~fb$^{-1}$ from proton-proton collisions collected with the LHCb
detector at centre-of-mass energies of 7\,TeV and 8\,TeV in 2011 and
2012, respectively. The isospin asymmetries are both consistent with
the Standard Model expectations. The three measured branching 
fractions favour lower values than their respective theoretical 
predictions, however they are all individually consistent with the Standard Model.

\end{abstract}

\vspace*{1.0cm}

\begin{center}
  Submitted to JHEP
\end{center}

\vspace{\fill}

{\footnotesize 
\centerline{\copyright~CERN on behalf of the \lhcb collaboration, license \href{http://creativecommons.org/licenses/by/3.0/}{CC-BY-3.0}.}}
\vspace*{2mm}

\end{titlepage}


\newpage
\setcounter{page}{2}
\mbox{~}
\newpage


\centerline{\large\bf LHCb collaboration}
\begin{flushleft}
\small
R.~Aaij$^{41}$, 
B.~Adeva$^{37}$, 
M.~Adinolfi$^{46}$, 
A.~Affolder$^{52}$, 
Z.~Ajaltouni$^{5}$, 
J.~Albrecht$^{9}$, 
F.~Alessio$^{38}$, 
M.~Alexander$^{51}$, 
S.~Ali$^{41}$, 
G.~Alkhazov$^{30}$
% ...

{\footnotesize \it
$ ^{1}$Centro Brasileiro de Pesquisas F\'{i}sicas (CBPF), Rio de Janeiro, Brazil\\
% ...
$ ^{5}$Clermont Universit\'{e}, Universit\'{e} Blaise Pascal, CNRS/IN2P3, LPC, Clermont-Ferrand, France\\
% ...
$ ^{9}$Fakult\"{a}t Physik, Technische Universit\"{a}t Dortmund, Dortmund, Germany\\
% ...
$ ^{30}$Petersburg Nuclear Physics Institute (PNPI), Gatchina, Russia\\
% ...
$ ^{37}$Universidad de Santiago de Compostela, Santiago de Compostela, Spain\\
$ ^{38}$European Organization for Nuclear Research (CERN), Geneva, Switzerland\\
% ...
$ ^{41}$Nikhef National Institute for Subatomic Physics, Amsterdam, The Netherlands\\
% ...
$ ^{46}$H.H. Wills Physics Laboratory, University of Bristol, Bristol, United Kingdom\\
% ...
$ ^{51}$School of Physics and Astronomy, University of Glasgow, Glasgow, United Kingdom\\
$ ^{52}$Oliver Lodge Laboratory, University of Liverpool, Liverpool, United Kingdom\\
% ...
}
\end{flushleft}

\cleardoublepage


\renewcommand{\thefootnote}{\arabic{footnote}}
\setcounter{footnote}{0}


\pagestyle{plain} 
\setcounter{page}{1}
\pagenumbering{arabic}


\def\ellell     {\ensuremath{\ell^+ \ell^-}\xspace}

\section{Introduction}

The rare decay of a \B meson into a strange meson and a \mumu pair is
a \bquark\to\squark quark-level transition. In the Standard Model~(SM),
this can only proceed via loop diagrams. The
loop-order suppression of the SM amplitudes increases the sensitivity
to new virtual particles that can influence the decay amplitude at a
similar level to the SM contribution. The branching fractions of
\AllModes decays are highly sensitive to contributions from vector or
axial-vector like particles predicted in extensions of the SM.
However, despite recent progress in lattice
calculations~\cite{kmumulattice,kstmumulattice}, theoretical
predictions of the decay rates suffer from relatively large
uncertainties in the $B\rightarrow K^{(*)}$ form factor calculations.

To maximise sensitivity, observables can be constructed from ratios or
asymmetries where the leading form factor uncertainties cancel. The
\CP-averaged isospin asymmetry (\AI) is such an observable. It is
defined as

\begin{equation} 
\begin{split}
\AI & = \frac{\Gamma(\NeutModes) -\Gamma(\CharModes)}{\Gamma(\NeutModes) +\Gamma(\CharModes)}\phantom{\AI}\phantom{,} \\ 
 & = \frac{\mathcal{B}(\NeutModes) - ({\tau_0}/{\tau_+})\cdot \mathcal{B}(\CharModes)}{\mathcal{B}(\NeutModes) + ({\tau_0}/{\tau_+})\cdot\mathcal{B}(\CharModes)},
\end{split}
\label{eq:AI}
\end{equation}
where $\Gamma(f)$ and $\mathcal{B}(f)$ are the partial width and
branching fraction of the $\B\to f$ decay and $\tau_0/\tau_+$ is the
ratio of the lifetimes of the \Bz and \Bp mesons\footnote{The
  inclusion of charge conjugated processes is implied throughout this
  paper.}. The decays in the isospin ratio differ only by the charge
of the light (spectator) quark in the \B meson. The SM prediction for
\AI is $\mathcal{O}(1\%)$ in the dimuon mass squared, \qq, region
below the \jpsi
resonance~\cite{isospintheory,Khodjamirian:2012rm,roman}.  There is no
precise prediction for \AI for the \qq region above the \jpsi
resonance, but it is expected to be even smaller than at low
\qq~\cite{roman}. As \qq approaches zero, the isospin asymmetry of
\BKstMuMu is expected to approach the same asymmetry as in
\decay{\B}{\Kstar\gamma} decays, which is measured to be
$(5\pm3)\%$~\cite{HFAG}.

Previously, \AI has been measured by the BaBar~\cite{babarnew},
Belle~\cite{belleiso} and LHCb~\cite{LHCb-PAPER-2012-011}
collaborations, where measurements for the \BKMuMu decay have
predominantly given negative values of \AI. In particular, the \BKMuMu
isospin asymmetry measured by the LHCb experiment deviates from zero
by over 4 standard deviations. For \BKstMuMu, measurements of \AI are
consistent with zero.

This paper describes a measurement of the isospin asymmetry in \BKMuMu
and \BKstMuMu decays based on data collected with the \lhcb detector,
corresponding to an integrated luminosity of 1\invfb recorded in 2011
at a centre-of-mass energy $\sqrt{s}=7$~TeV, and 2\invfb recorded in
2012 at $\sqrt{s}=8$~TeV. The previous
analysis~\cite{LHCb-PAPER-2012-011} was carried out on the 1\invfb of
data recorded in 2011. The analysis presented here includes, in addition to
the data from 2012, a re-analysis of the full 1\invfb data sample with
improved detector alignment parameters, reconstruction algorithms and
event selection. Thus it supersedes the measurements in
Ref.~\cite{LHCb-PAPER-2012-011}. Moreover, the assumption that there
is no isospin asymmetry in the \JpsiModes decays is now used for all
the measurements.

The isospin asymmetries are determined by measuring the differential
branching fractions of \mbox{\BuKMuMu}, \mbox{\BdKzMuMu},
\mbox{\BdKstarzMuMu} and \mbox{\BuKstarpMuMu} decays. The \Kz meson is
reconstructed through the decay $\KS\rightarrow\pi^+\pi^-$; the
\Kstarp as $\Kstarp\rightarrow \KS(\rightarrow\pi^+\pi^-)\pi^+$ and
the $\Kstarz$ as $\Kstarz\rightarrow K^+\pi^-$. Modes involving a \KL
or \piz in the final state are not considered. The individual
branching fractions of \mbox{\BuKMuMu}, \mbox{\BdKzMuMu} and
\mbox{\BuKstarpMuMu} decays are also reported. The branching fraction
of the decay \BdKstMuMu has been previously reported in
Ref.~\cite{LHCb-PAPER-2013-019} and is not updated here.

The \BdKstMuMu and \BuKstMuMu branching fractions are influenced by
the presence of \decay{\Bz}{\Kp\pim\mumu} and
\decay{\Bp}{\KS\pip\mumu} decays with the $\Kp\pim$ or $\KS\pip$
system in a S-wave configuration. It is not possible to separate these
candidates from the dominant \Kstarz and \Kstarp resonant components without performing an analysis
of the $\Kp\pim$ or $\KS\pip$ invariant mass and the angular
distribution of the final state particles. The S-wave component is
expected to be at the level of a few percent~\cite{Becirevic:2012dp}
and to cancel when evaluating the isospin asymmetry of the \BKstMuMu
decays.

\section{Detector and dataset}
\label{sec:Normalisation}

The \lhcb detector~\cite{Alves:2008zz} is a single-arm forward
spectrometer covering the \mbox{pseudorapidity} range $2<\eta <5$,
designed for the study of particles containing \bquark or \cquark
quarks. The detector includes a high-precision tracking system
consisting of a silicon-strip vertex detector surrounding the $pp$
interaction region, a large-area silicon-strip detector located
upstream of a dipole magnet with a bending power of about
$4{\rm\,Tm}$, and three stations of silicon-strip detectors and straw
drift tubes~\cite{LHCb-DP-2013-003} placed downstream of the magnet.
The combined tracking system provides a momentum measurement with
relative uncertainty that varies from 0.4\% at 5\gevc to 0.6\% at
100\gevc, and an impact parameter resolution of 20\mum for tracks with
high transverse momentum. Charged hadrons are identified using two
ring-imaging Cherenkov~(RICH)
detectors~\cite{LHCb-DP-2013-001}. Photon, electron and hadron
candidates are identified by a calorimeter system consisting of
scintillating-pad and preshower detectors, an electromagnetic
calorimeter and a hadronic calorimeter. Muons are identified by a
system composed of alternating layers of iron and multiwire
proportional chambers~\cite{LHCb-DP-2012-003}.  Decays of
\decay{\KS}{\pip\pim} are reconstructed in two different categories:
the first involving \KS mesons that decay early enough for the
daughter pions to be reconstructed in the vertex detector; and the
second containing \KS mesons that decay later such that track segments
of the pions cannot be formed in the vertex detector. These categories
are referred to as \emph{long} and \emph{downstream},
respectively. Candidates in the long category have better mass,
momentum and vertex resolution than those in the downstream category.

Simulated events are used to estimate the efficiencies of the trigger,
reconstruction and subsequent event selection of the different signal
decays and to estimate the contribution from specific background
sources. These samples are produced using the software described in
Refs.~\cite{Sjostrand:2006za,Sjostrand:2007gs,LHCb-PROC-2010-056,Lange:2001uf,Golonka:2005pn,Allison:2006ve,*Agostinelli:2002hh,LHCb-PROC-2011-006}.

\section{Selection}
\label{sec:Selection}
The \AllModes candidate events are required to pass a two-stage
trigger system~\cite{LHCb-DP-2012-004}. In the initial hardware
stage, these events are selected with at least one muon with transverse
momentum, $\pt>1.48\,(1.76)\gevc$ in 2011~(2012). In the subsequent
software stage, at least one of the final-state particles is required
to have $\pt> 1.0\gevc$ and an impact parameter~(IP) larger than
$100\mum$ with respect to all of the primary $pp$ interaction
vertices~(PVs) in the event. Finally, a multivariate
algorithm~\cite{BBDT} is used for the identification of secondary
vertices consistent with the decay of a \bquark hadron with muons in
the final state.

For the \mbox{\BdKsMuMu} and \mbox{\BuKstarpMuMu} modes, \KS
candidates are required to have a mass within 30~\mevcc of the known
\KS mass~\cite{PDG2012}. For the \mbox{\BdKstarzMuMu} and
\mbox{\BuKstarpMuMu} modes, \Kstar candidates are formed by combining
kaons and pions and are required to have a mass within 100\mevcc of
the known \Kstar masses~\cite{PDG2012}. For all decay modes, \B candidates
are formed by subsequently combining the $K^{(*)}$ meson with two
muons of opposite charge and requiring the mass to be between 5170 and
5700\mevcc.

The event selection is common to that described in
Refs.~\cite{LHCb-PAPER-2013-039,LHCb-PAPER-2013-019,angular_paper}: the
$\mu^\pm$ and the \Kp candidates are required to have $\chi^{2}_{\rm IP} > 9$, where $\chi^{2}_{\rm IP}$ is
defined as the minimum change in $\chi^2$ of the vertex fit to any of
the PVs in the event when the particle is added to that PV; the dimuon
pair vertex fit has $\chisq<9$; the \B candidate
is required to have a vertex fit $\chisq < 8$ per degree of freedom; 
the \B momentum vector is aligned with respect to one of the PVs in the event 
within 14\mrad, the \B candidate has $\chi^{2}_{\rm IP} < 9$ with respect to that
PV and the vertex fit \chisq of that PV increases by more than 121 when 
including the \B decay products. In addition, the \KS candidate is
required to have a decay time larger than 2\ps. 

Using particle identification information from the RICH detectors,
calorimeters and muon system, multivariate discriminants~(PID
variables) are employed to reject background candidates, where pions
are misidentified as kaons and vice-versa, and where a pion or kaon is
incorrectly identified as a muon.

The initial selection is followed by a tighter multivariate selection,
based on a boosted decision tree~(BDT)~\cite{Breiman} with the
AdaBoost algorithm\cite{AdaBoost}, which is designed to reject
background of combinatorial nature. Separate BDTs are employed for
each signal decay. For decays involving a \KS meson, two
independent BDTs are trained for the long and downstream categories.
This gives a total of six BDTs which all use data from the upper
mass sideband ($m(K^{(*)}\mumu)>5350$~\mevcc) of their corresponding
decay to represent the background sample in the training. Simulated
\mbox{\BuKMuMu}, \mbox{\BdKsMuMu} and \mbox{\BuKstarpMuMu} events are
used as the signal sample in the training of the corresponding
BDTs. In contrast, to stay consistent with the selection in
Ref.~\cite{LHCb-PAPER-2013-039}, the signal for the training of the
\mbox{\BdKstarzMuMu} BDT is taken from reconstructed
\mbox{\BdJpsimumuKstarz} candidates from data.  Events used in the
training of the BDTs are not used in the subsequent classification of
the data.

All six BDTs use predominantly geometric variables, including the
variables used in the pre-selection described above. The
\mbox{\BdKstarzMuMu} BDT also makes use of PID variables to further
suppress background where a \Kp is misidentified as a \pip and
vice-versa in the \Kstarz decay.

The multivariate selections for \mbox{\BuKMuMu}, \mbox{\BuKstarpMuMu}
and \mbox{\BdKstarzMuMu} candidates have an efficiency of $90\%$ for
the signal channels and remove $95\%$ of the background that remains
after the pre-selection. The long lifetime of the \KS
meson makes it difficult to determine whether it originates from the
same vertex as the dimuon system in \mbox{\BdKsMuMu} decays. As such,
the multivariate selection for \mbox{\BdKsMuMu} candidates has a
signal efficiency of 66\% and 48\% for the long and downstream
categories, respectively, while removing 99\% of the background
surviving the pre-selection.

Combinatorial background, where the final-state particles attributed to the \B
candidate do not all come from the same \bquark-hadron decay, are
reduced to a small level by the multivariate selection. In addition,
there are several sources of background that peak in the
$K^{(*)}\mumu$ invariant mass. The largest of these are
\decay{\B}{\jpsi K^{(*)}} and \decay{\B}{\psitwos K^{(*)}} decays,
which are rejected by removing the regions of dimuon invariant mass
around the charmonium resonances $(2828 < m({\mumu}) < 3317\mevcc$ and
$3536 < m({\mumu}) < 3873\mevcc)$. A combination of mass and PID
requirements remove additional peaking backgrounds. These include
\decay{\Lb}{\PLambda^{(*)}\mumu} decays, where the proton from the
\decay{\PLambda}{p\pim} decay is misidentified as a \Kp or the proton
misidentified as a \pip in the \decay{\PLambda^{*}}{p\Km} decay,
\decay{\Bs}{\phi\mumu} decays where a kaon from \decay{\phi}{\Kp\Km}
is misidentified as a pion, and \BuKMuMu decays that combine with a
random pion to fake a \BdKstarzMuMu decay. After the application of
all the selection criteria the exclusive backgrounds are reduced to
less than $1\%$ of the level of the signal.

To improve the resolution on the reconstructed mass of the
$B$ meson, a kinematic fit~\cite{Hulsbergen:2005pu} is performed for
candidates involving a \KS meson. In the fit, the mass of the
$\pi^+\pi^-$ system is constrained to the nominal \KS mass and the $B$
candidate is required to originate from its associated PV.

\section{Signal yield determination}
\label{sec:massfits}

\begin{figure}[t]
\centering
\subfigure{\includegraphics[width=0.45\textwidth] {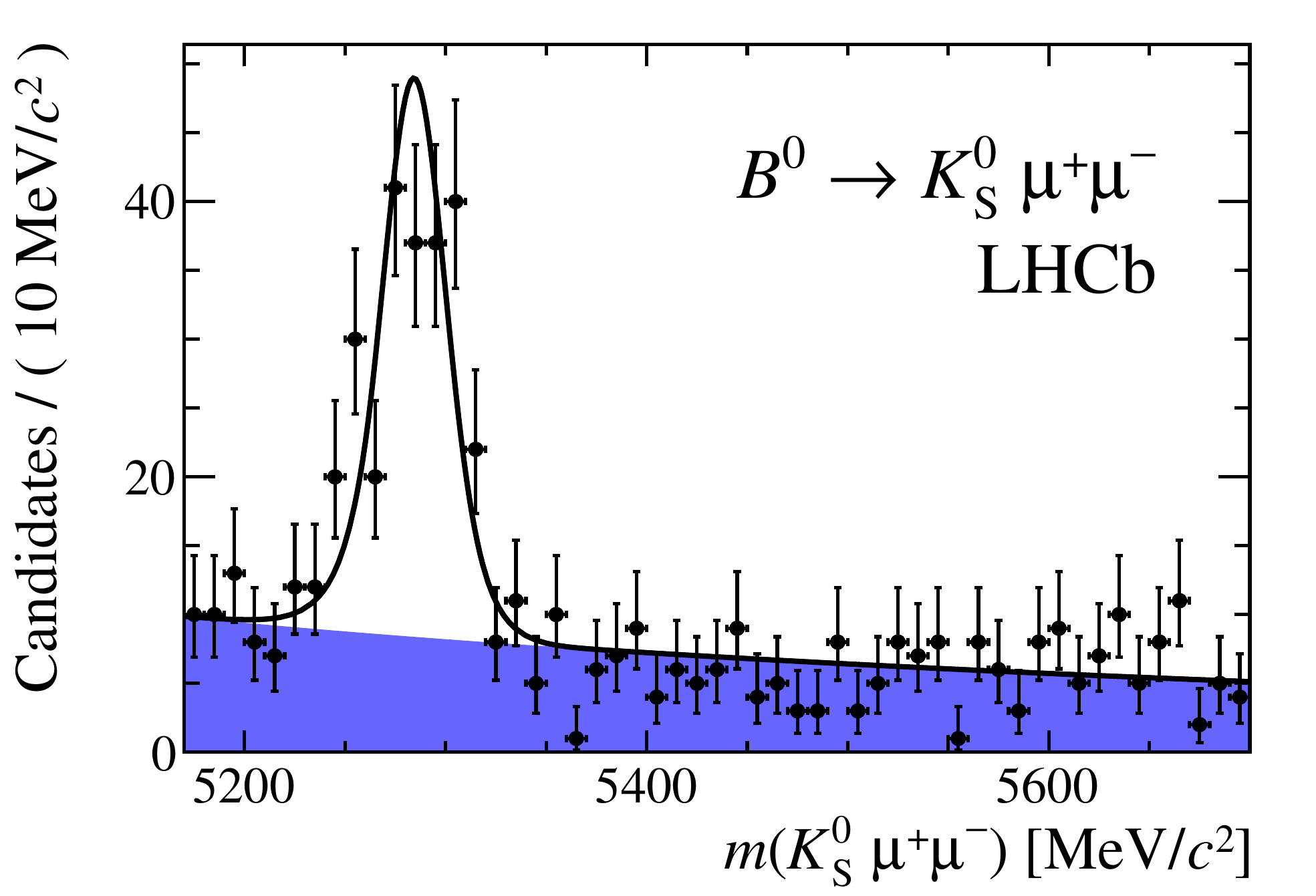}}
\subfigure{\includegraphics[width=0.45\textwidth] {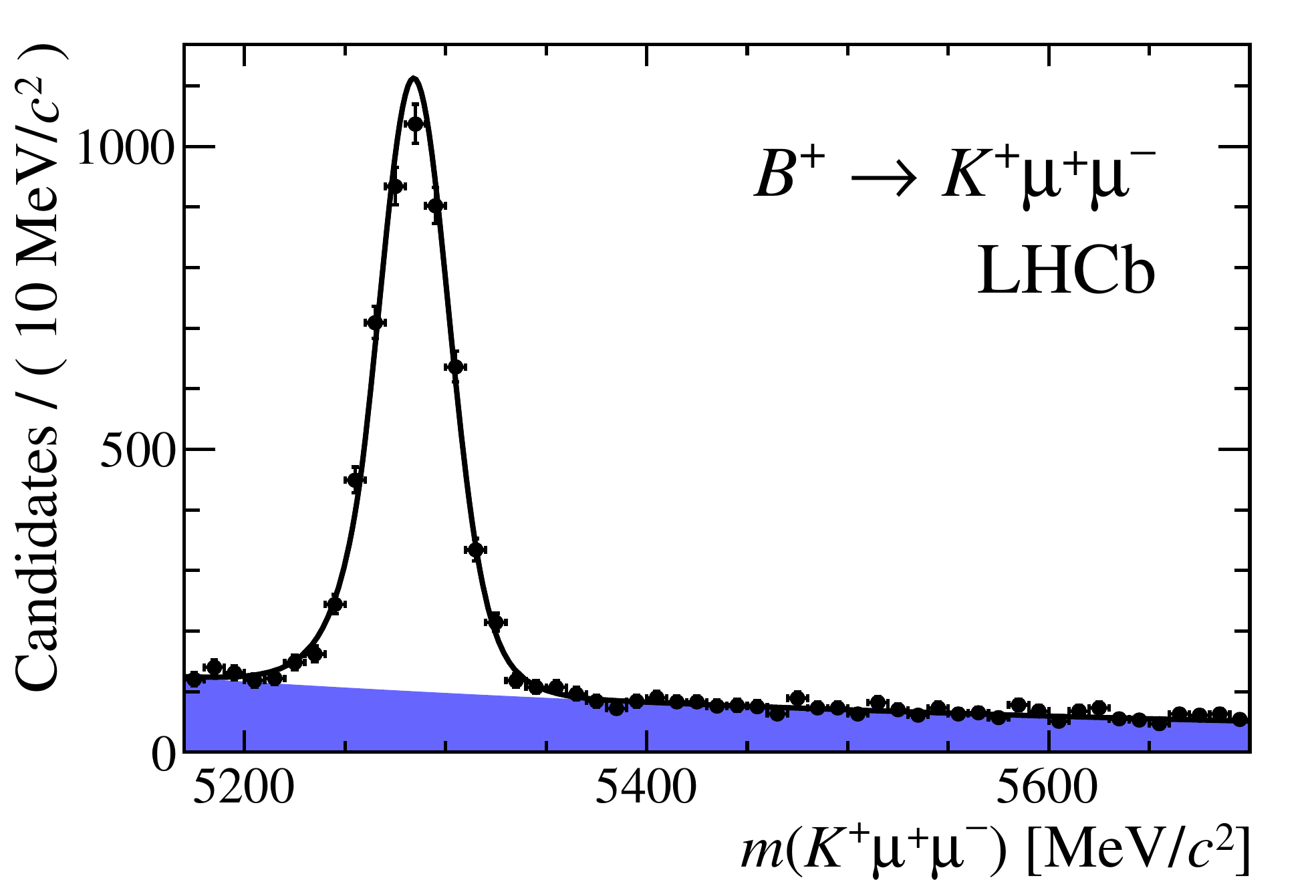}}
\subfigure{\includegraphics[width=0.45\textwidth] {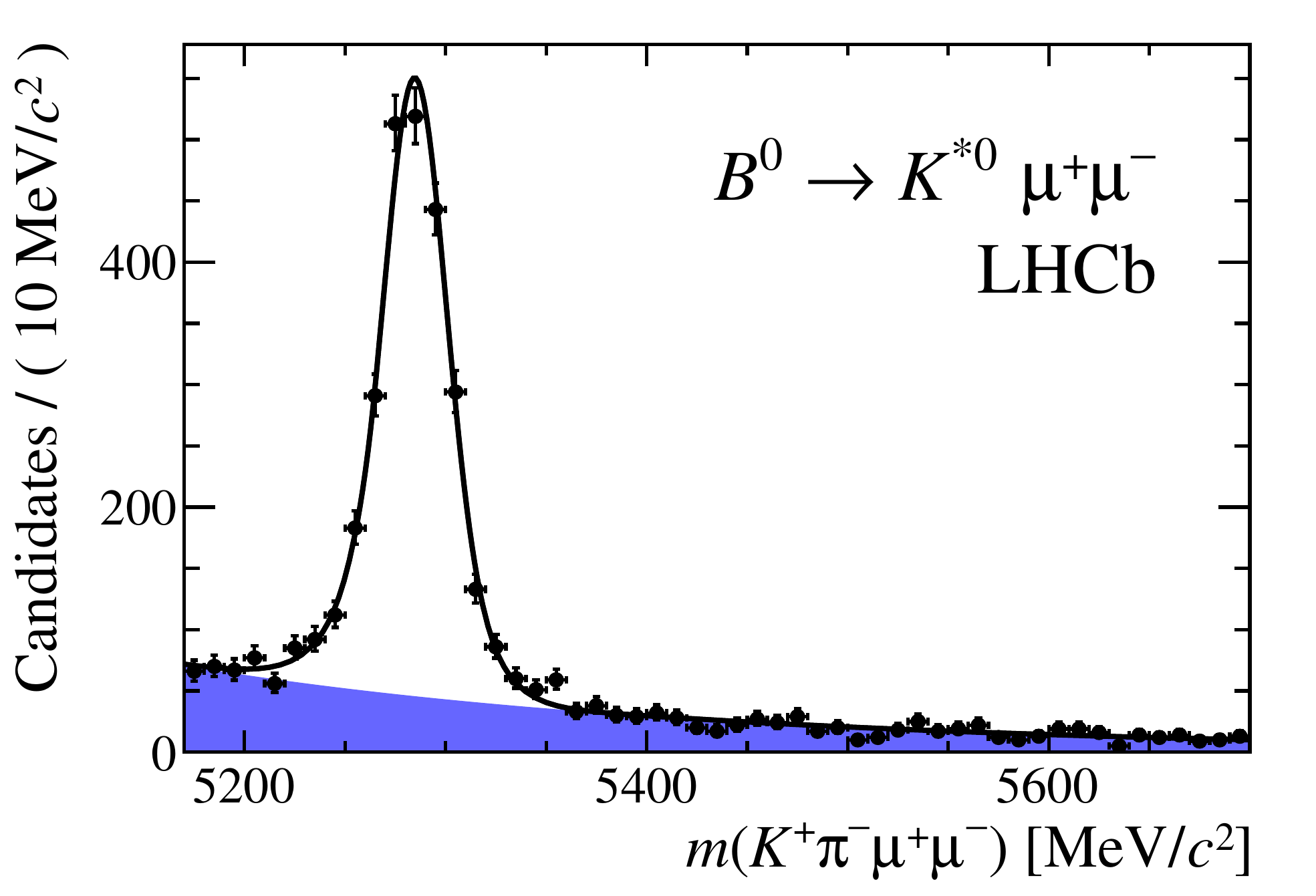}}
\subfigure{\includegraphics[width=0.45\textwidth] {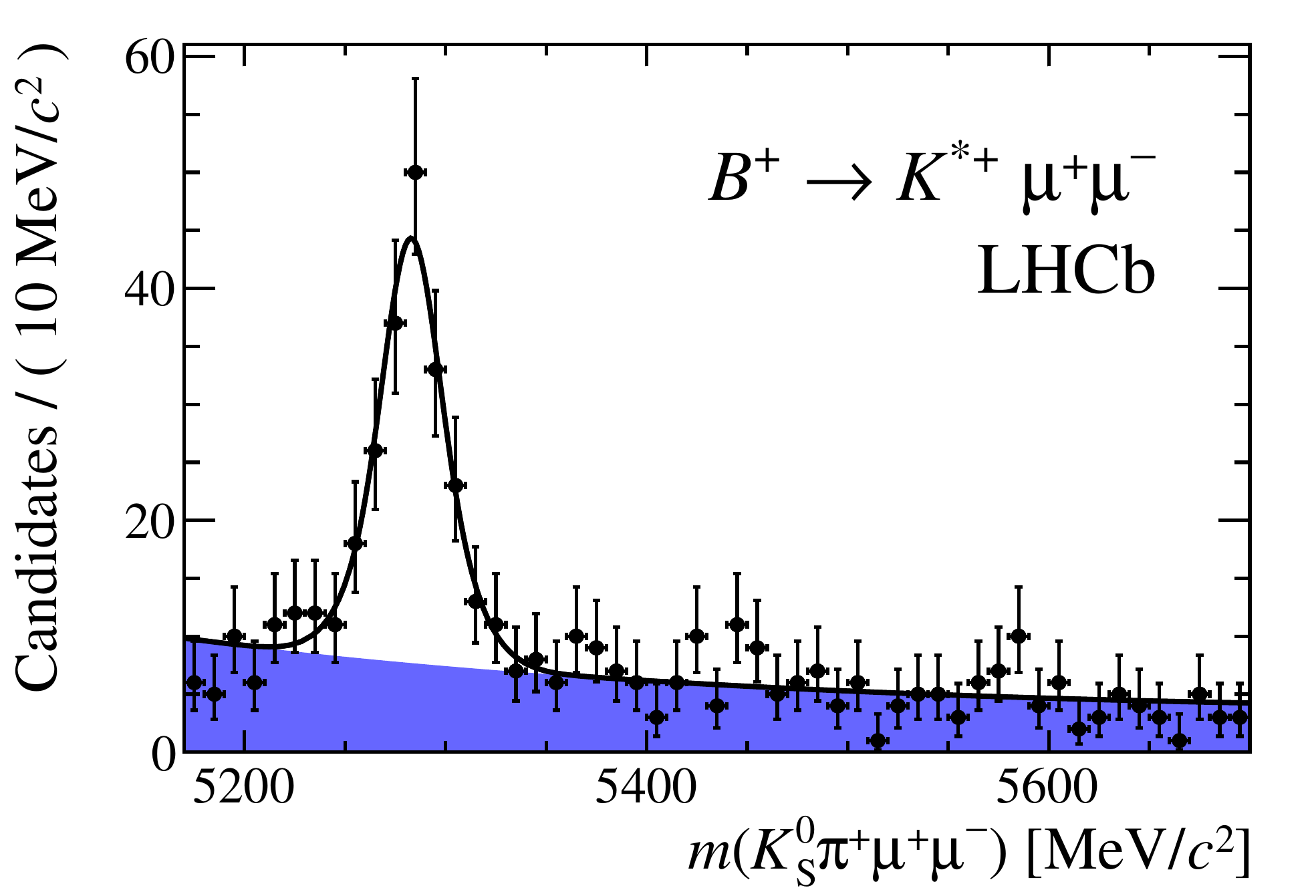}}
\caption{\small Reconstructed \B candidate mass for the four signal modes.
  The data are overlaid with the result of the fit described in the
  text. The long and downstream \KS categories are combined.
  The results of the fits, performed in separate \qq bins, are merged
  for presentation purposes. The blue~(shaded) region is the
  combinatorial background.}
  \label{fig:mass}
\end{figure}

Signal yields are determined using extended unbinned maximum
likelihood fits to the $\kaon^{(*)}\mumu$ mass in the range
5170--5700\mevcc. These fits are performed in nine bins of \qq for
\mbox{\BdKsMuMu}, \mbox{\BuKstarpMuMu} and \mbox{\BdKstMuMu} decays,
while for the \mbox{\BuKMuMu} decay the larger number of signal events allows to
define nineteen \qq bins. The binning scheme is shown in
Tables~\ref{tab:kmumu_results} to~\ref{tab:bukst_results} of the
appendix. It removes the region of \qq around the charmonium
resonances. For the \BuKMuMu differential branching fraction, where
the statistical uncertainty is the smallest, a narrow range in
$m(\mumu)$ is also removed around the mass of the $\phi$ meson. The
signal component in the fit is described by the sum of two Crystal
Ball functions~\cite{Skwarnicki:1986xj} with common peak values and
tail parameters, but different widths. The signal shape parameters are
taken from a fit to \JpsiModes channels in the data, with a correction
that accounts for a small \qq dependence on the peak value and width
obtained from the simulation. The combinatorial background is
parameterised by an exponential function, which is allowed to vary for
each \qq bin and \KS category independently. For decays involving \KS
mesons, separate fits are made to the long and downstream
categories. The mass fits for the four signal channels are shown in
Fig.~\ref{fig:mass}, where the long and downstream \KS categories are
combined and the results of the fits, performed in separate \qq bins,
are merged for presentation purposes. The corresponding number
of signal candidates for each channel is given in
Table~\ref{tab:signal_yields}.
\begin{table}
  \centering
  \caption{\small Observed yields of the four signal channels
    summed over the \qq bins, excluding the charmonium resonance
    regions. Only the statistical uncertainties are shown.}
  \setlength{\extrarowheight}{2pt}
  \begin{tabular}{l r}
    \hline
     Decay mode & Signal yield \\
    \hline
    \BuKMuMu     & $4746 \pm 81$ \\
    \BdKsMuMu  & $176 \pm 17$ \\
    \decay{\Bp}{\Kstarp (\to \KS\pip)\mumu} & $162\pm16$\\
    \decay{\Bz}{\Kstarz (\to \Kp\pim)\mumu} & $2361\pm56$ \\
    \hline
  \end{tabular}
  \label{tab:signal_yields}
\end{table}
 
\section{Branching fraction normalisation}
\label{sec:Normalisation}

Each signal mode is normalised with respect to its corresponding
\JpsiModes channel, where the \jpsi resonance decays into two
muons. These normalisation channels have branching fractions that are
approximately two orders of magnitude higher than those of the signal
channels. Each normalisation channel has similar kinematic properties
and the same final-state particles as the signal modes. This results
in an almost complete cancellation of systematic uncertainties when
measuring the ratio of branching fractions of the signal mode with the
corresponding normalisation channel. Separate normalisations for the
long and downstream \KS reconstruction categories are used to further
cancel potential sources of systematic uncertainty.

Corrections to the IP resolution, PID variables and \B candidate
kinematic properties are applied to the simulated events, such that the
distributions of simulated candidates from the normalisation channels
agree with the data. The simulation samples are subsequently used to
calculate the relative efficiencies as functions of \qq. The \qq
dependence arises mainly from trigger effects, where the muons have
increased (decreased) \pt at high (low) \qq and consequently have a
higher (lower) trigger efficiency.  Furthermore, at high \qq, the hadrons
are almost at rest in the \B meson rest frame and, like the \B meson, points
back to the PV in the laboratory frame.  The IP requirements applied
on the hadron have a lower efficiency for this
region of \qq. The \KS channels have an additional effect due to the
different acceptance of the two reconstruction categories; \KS mesons
are more likely to be reconstructed in the long category if they have
low momentum, which favours the high \qq region. The momentum
distributions of the \KS mesons in \BdJpsiKs and \BuJpsiKst decays in data
and simulation for both \KS categories are in good agreement,
indicating that the acceptance is well described in the simulation.

The measured differential branching fraction averaged over a \qq bin of width
$q^2_{\rm max}-q^2_{\rm min}$ is given by
\begin{equation} 
\frac{d\mathcal{B}}{dq^{2}} = \frac{N(\AllModes)}{N(\JpsiModes)}\cdot 
\frac{\varepsilon(\JpsiModes)}{\varepsilon(\AllModes)}\cdot
\frac{\mathcal{B}(\JpsiModes)\mathcal{B}(\jpsi\to\mup\mun)}{(q^{2}_{\rm max} - q^{2}_{\rm min})},
\label{eq:BF}
\end{equation}
where ${N(\AllModes})$ is the number of signal candidates in
the bin, ${N(\JpsiModes})$ is the number of normalisation candidates, the
product of \mbox{$\mathcal{B}(\JpsiModes)$} and
\mbox{$\mathcal{B}(\jpsi\to\mup\mun)$} is the visible branching
fraction of the normalisation channel, and
${\varepsilon(\AllModes)}/{\varepsilon(\JpsiModes)}$ is the relative
efficiency between the signal and normalisation channels in the bin.

\section{Systematic uncertainties}
\label{sec:Systematics}

The branching fraction measurements of the normalisation modes from
the $B$-factory experiments assume that the \Bu and \Bd mesons are
produced with equal proportions at the $\Upsilon(4S)$
resonance~\cite{Aubert:2004rz,Abe:2002rc,Abe:2002haa}. In contrast, in
this paper isospin symmetry is assumed for the \decay{\B}{\jpsi
  K^{(*)}} decays, implying that the \BuJpsiK (\BuJpsiKst) and \BdJpsiKz
(\BdJpsiKst) decays have the same partial width. The branching fractions
 used in the normalisation are obtained by: taking the most
precise branching fraction results from Ref.~\cite{Aubert:2004rz} and
translating them into partial widths; averaging the partial widths of the
\Kp, \Kz and the \Kstarp, \Kstarz modes, respectively; and finally
translating the widths back to branching fractions. The calculation only
requires knowledge of the ratio of \Bz and \Bp lifetimes for which we
use $0.93\pm0.01$~\cite{PDG2012}. Statistical uncertainties are
treated as uncorrelated while systematical uncertainties are
conservatively treated as fully correlated. The resulting branching
fractions of the normalisation channels are
\begin{displaymath}
\begin{split}
\mathcal{B}(\BuJpsiK) & = ( 0.998  \pm 0.014 \pm 0.040 ) \times 10^{-3}, \\
\mathcal{B}(\BdJpsiKz) & = ( 0.928 \pm 0.013 \pm 0.037 ) \times 10^{-3},\\
\mathcal{B}(\decay{\Bp}{\jpsi \Kstarp}) & = ( 1.431 \pm 0.027 \pm 0.090 ) \times 10^{-3},\\
\mathcal{B}(\BdJpsiKst) & = ( 1.331 \pm 0.025 \pm 0.084 ) \times 10^{-3}\nonumber,
\end{split}
\label{eq:JpsiBFs}
\end{displaymath}
where the first uncertainty is statistical and the second
systematic.

A systematic uncertainty is assigned to account for the imperfect
knowledge of the \qq spectrum in the simulation within each \qq
bin. For example, the recent observation of a resonance in the high
\qq region of \BuKMuMu decays~\cite{LHCb-PAPER-2013-039} alters the
\qq distribution and hence the selection efficiencies in that
region. By reweighting simulated events to account for this resonance,
and for variations of the $B\rightarrow K^{(*)}$ form factor model as described in
 Ref.~\cite{Ball:2004ye}, a
systematic uncertainty is determined at the level of $(1-2)\%$ depending
on channel and \qq bin.

Data-driven corrections of the long and downstream tracking efficiencies in
 the simulation are determined using tag-and-probe techniques in 
 $J/\psi\rightarrow\mu^+\mu^-$ and $\Dz\to\phi\KS$ decays, respectively. For the $J/\psi\rightarrow\mu^+\mu^-$ decay, 
 the tag is a fully reconstructed muon track. It is combined with another muon,
  referred to as the probe, reconstructed using the muon stations and the large-area silicon detector
   upstream of the magnet. The tracking efficiency is determined by reconstructing
    the probe using the full tracking system. The $\Dz\to\phi\KS$  decay is tagged 
    via a partial reconstruction using only one of the \KS daughters. The 
    downstream tracking efficiency is then evaluated by fully reconstructing the \KS candidate. 
The resulting systematic uncertainty on the efficiency ratio, due to finite precision of the
measurement, is found to be negligible. The systematic uncertainty
that arises from the corrections to the IP resolution, PID variables
and \B candidate kinematic properties in the simulation varies between
1\% and 3\% depending on channel and \qq bin.

A summary of the systematic uncertainties can be found in Table~\ref{tab:syst}. 
The uncertainties on the branching fractions of the
normalisation modes constitute the dominant source of systematic
uncertainty on the branching fraction measurements while it cancels in
the isospin measurements.

\begin{table}[ht]
  \centering
  \caption{Summary of systematic uncertainties associated with the branching 
  fraction and isospin asymmetry measurements.}
    \begin{tabular}{l c c}
      \hline
      \noalign{\vskip 1mm}
      Source & Branching fraction & Isospin asymmetry  \\
      \hline
      \noalign{\vskip 1mm}
\JpsiModes branching fractions & $4\%-6$\% & $-$ \\ 
Physics model & $1\%-2$\% & $1\%-2$\% \\
Simulation mis-modelling & $1\%-3$\% & $1\%-3$\% \\
      \hline
    \end{tabular}
  \label{tab:syst}
  \end{table}


\section{Branching fraction results}
\label{sec:BrResults}

\begin{figure}[tb]
\centering
\subfigure{\includegraphics[width=0.45\textwidth] {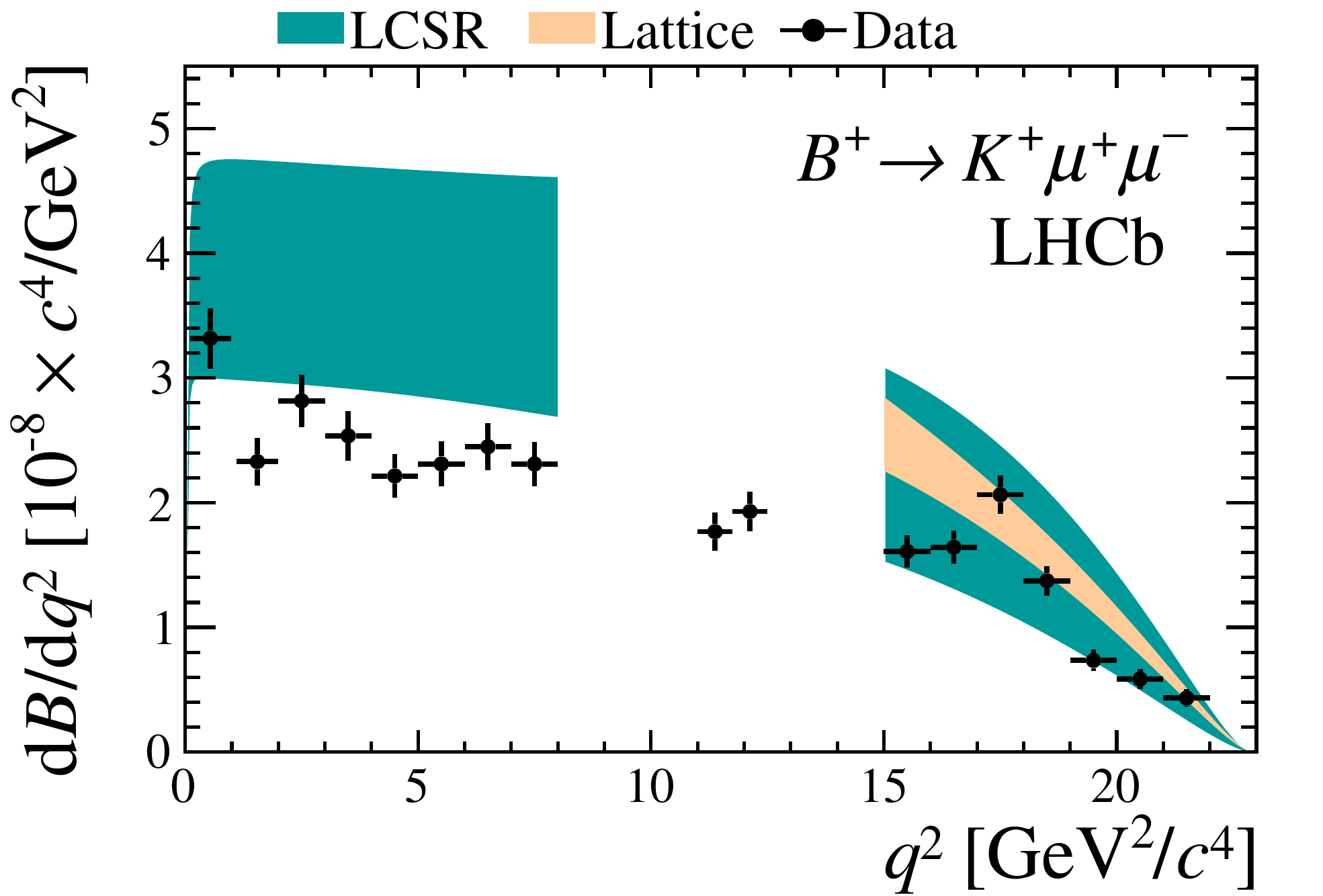}}
\subfigure{\includegraphics[width=0.45\textwidth] {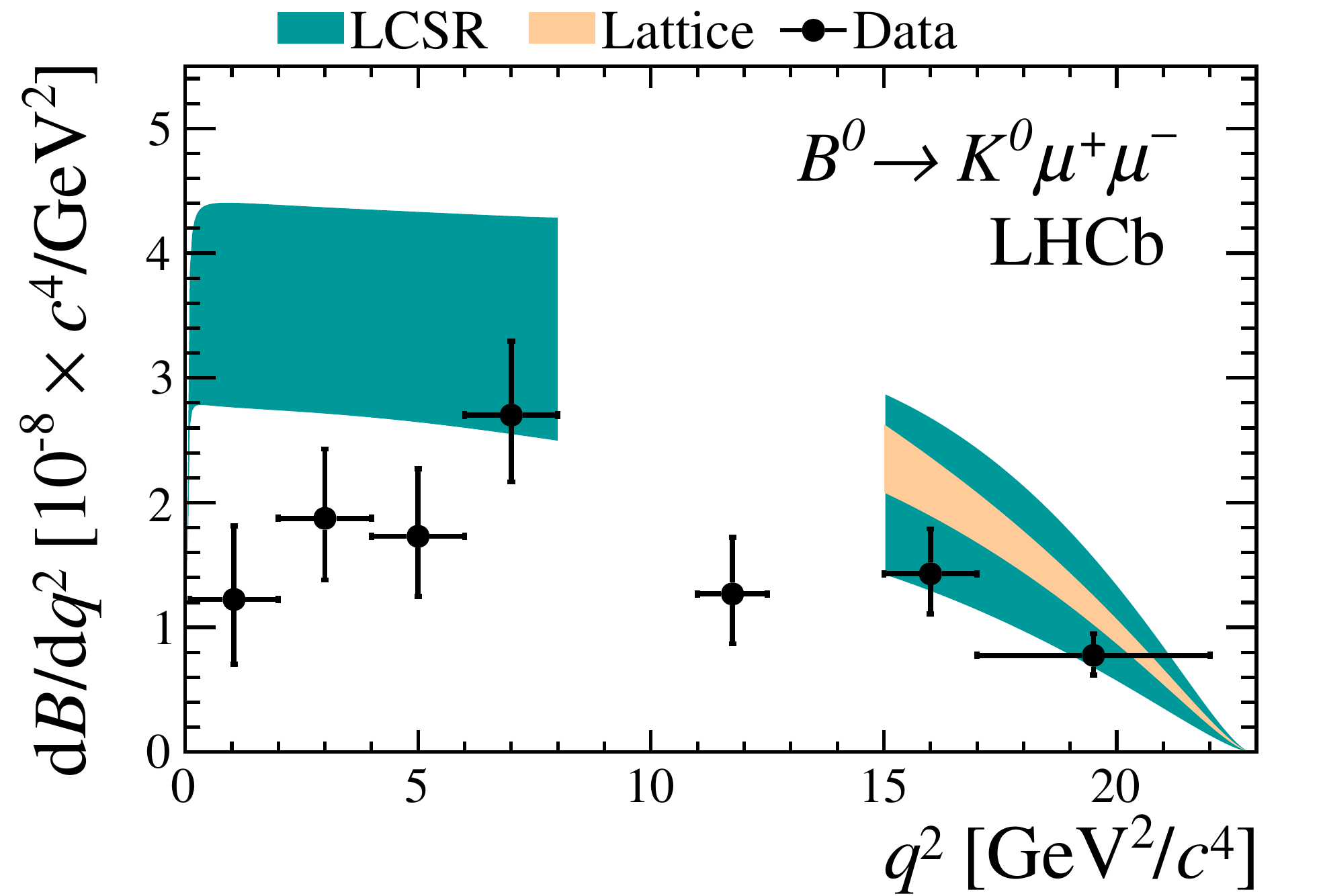}}
\subfigure{\includegraphics[width=0.45\textwidth] {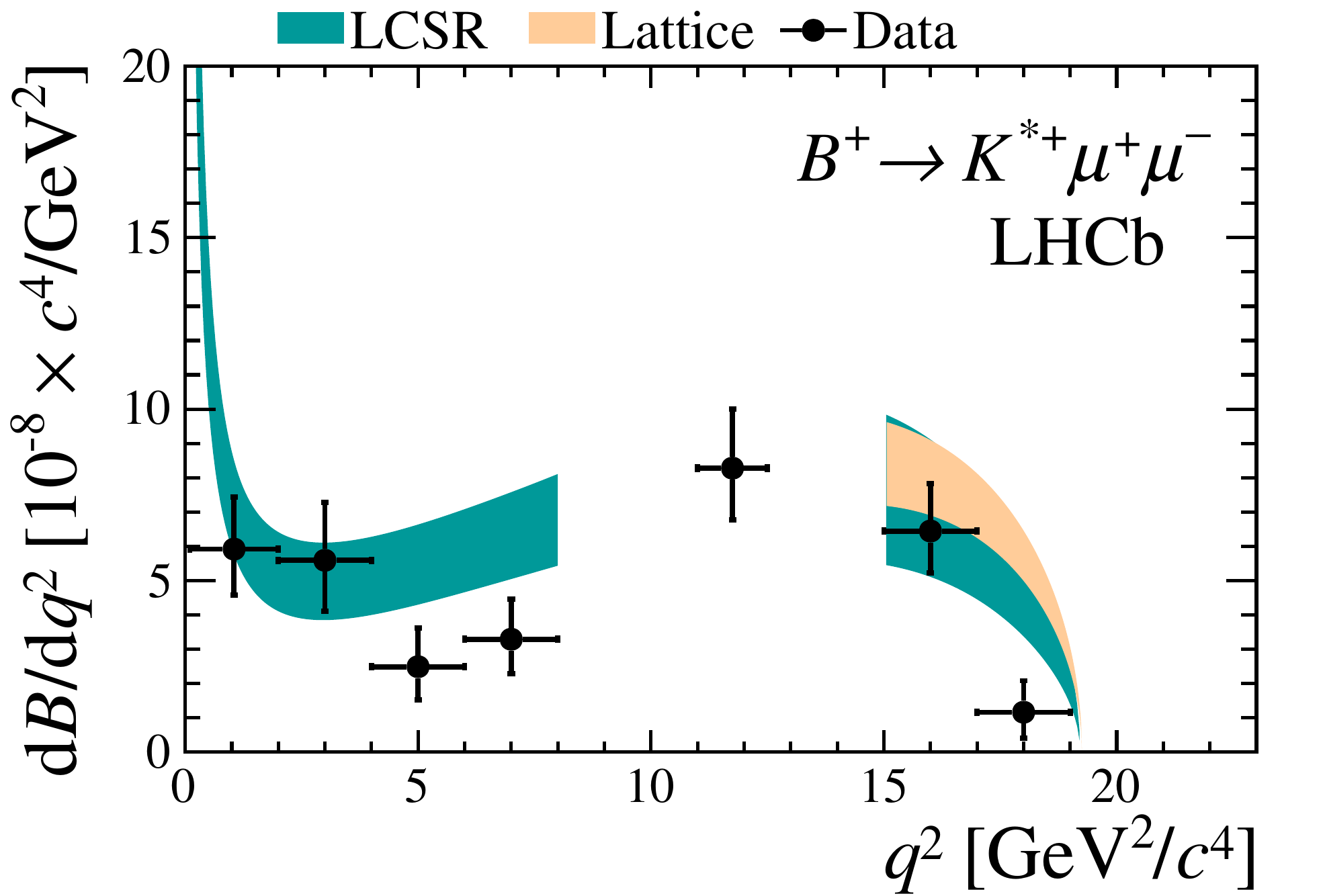}}
\caption{\small Differential branching fraction results for the
  \BuKMuMu, \BdKzMuMu and \BuKstMuMu decays. The uncertainties shown on
  the data points are the quadratic sum of the statistical and
  systematic uncertainties. The shaded regions illustrate the
  theoretical predictions and their uncertainties from light cone sum
  rule and lattice QCD calculations.}
  \label{fig:kmumu_BFs}
\end{figure}

The differential branching fraction results for \BuKMuMu, \BdKzMuMu
and \BuKstMuMu decays are shown in Fig.~\ref{fig:kmumu_BFs} with
theoretical predictions~\cite{Bobeth:2011gi,Bobeth:2011nj}
superimposed. The values are given in Tables~\ref{tab:kmumu_results}
to~\ref{tab:bukst_results} in the appendix. In the low \qsq region,
these predictions rely on the QCD factorisation approaches from
Refs.~\cite{Beneke:2001at,Beneke:2004dp} for \BKstMuMu and
Ref.~\cite{Bobeth:2007dw} for \BKMuMu, and lose accuracy when
approaching the \jpsi resonance. In the high \qsq region, an operator
product expansion in the inverse $b$-quark mass, $1/m_\bquark$, and in
$1/\sqrt{\qsq}$ is used based on Ref.~\cite{Grinstein:2004vb}. This
expansion is only valid above the open charm threshold. A dimensional
estimate of the uncertainty associated with this expansion is
discussed in Ref.~\cite{Egede:2008uy}. For light cone sum rule~(LCSR)
predictions, the $B\rightarrow K^{(*)}$ form factor calculations are
taken from Refs.~\cite{Ball:2004rg}
and~\cite{Khodjamirian:2010vf}. Predictions based on form factors from
lattice calculations are also
overlaid~\cite{kmumulattice,kmumulattice2,kstmumulattice,kstmumulattice2}.

Although all three differential branching fraction measurements are
consistent with the SM, they all have values smaller than the
theoretical prediction. The sample size for \BuKMuMu is sufficient to
show significant structures in the \qq distribution. As an
example, the peak at high \qq is due to the $\psi(4160)$ resonance,
which is discussed in more detail in Ref.~\cite{LHCb-PAPER-2013-039}.

The presence of an S-wave contribution to the $K^+\pi^-$ and
$\KS\pi^+$ systems of \BdKstarzMuMu and \BuKstMuMu candidates,
respectively, complicates the analysis of these channels. This effect
is of the order of a few percent and can be neglected in
\BuKstMuMu decays with the current statistical precision. The larger
signal yield of \BdKstarzMuMu, however, merits a detailed analysis of
the S-wave contribution and requires a dedicated study. For this reason the
branching fraction of \BdKstarzMuMu decays is not reported.

By convention, branching fractions are extrapolated to the full \qq
range ignoring the presence of the narrow charmonium resonances. A
\qq distribution based on Ref.~\cite{Ali:1999mm} is used for this. The
correction factors to the branching fractions due to this
extrapolation are 1.39 and 1.50 for \BKMuMu and \BdKstarzMuMu,
respectively. No uncertainty is assigned to these corrections.
Summing the \qq bins and applying the extrapolation, the
integrated branching fractions become
  \begin{equation}
  \begin{split}
      \mathcal{B}(\BuKMuMu) & = (4.29\pm0.07\,(\rm{stat})\pm0.21\,(\rm{syst}))\times 10^{-7}, \phantom{,} \\ 
      \mathcal{B}(\BdKzMuMu) & = (3.27\pm0.34\,(\rm{stat})\pm0.17\,(\rm{syst}))\times 10^{-7}, \phantom{,} \\ 
      \mathcal{B}(\BuKstMuMu) & = (9.24\pm0.93\,(\rm{stat})\pm0.67\,(\rm{syst}))\times 10^{-7}. \phantom{and} \nonumber
  \end{split}
    \end{equation}
These measurements are more precise than the current world
averages~\cite{PDG2012}.

Table~\ref{tab:lowrecoilBFs} compares the \BuKMuMu and \BdKzMuMu
branching fractions integrated over the \qq region of
$15-22\gevgevcccc$, and the \BuKstarpMuMu branching fraction
integrated over the $15-19\gevgevcccc$ region to the lattice QCD
predictions~\cite{kmumulattice,kmumulattice2,kstmumulattice,kstmumulattice2}.
While the measurements are all individually consistent with their
respective predictions, they all have values below those.
\begin{table}[t]
  \centering
  \caption{\small Integrated branching fractions $(10^{-8})$ in the high \qq 
    region. For the \BKMuMu modes the region is defined as $15-22\gevgevcccc$, 
    while for \BuKstarpMuMu it is $15-19\gevgevcccc$. Predictions are obtained 
    using the form factors calculated in lattice QCD over the same \qq 
    regions. For the measurements, the first uncertainty is statistical 
    and the second systematic.}
  \setlength{\extrarowheight}{3pt}
  \begin{tabular}{c c c}
    \hline
    Decay mode & Measurement & Prediction \\
    \hline
    \BuKMuMu & $\phantom{0}8.5\pm0.3\pm0.4$ & $10.7\pm1.2$ \\
    \BdKzMuMu & $\phantom{0}6.7\pm1.1\pm0.4$ & $\phantom{0}9.8\pm1.0$ \\
    \BuKstMuMu & $15.8~^{+3.2}_{-2.9}\pm1.1$ & $26.8\pm3.6$ \\
    \noalign{\vskip 1mm}
    \hline
  \end{tabular}
  \label{tab:lowrecoilBFs}
\end{table}

\section{Isospin asymmetry results}
\label{sec:AIResults}
The assumption of no isospin asymmetry in the \JpsiModes modes makes
the isospin measurement equivalent to measuring the difference in
isospin asymmetry between \AllModes and \JpsiModes decays. Compared to
using the values in Ref.~\cite{PDG2012} for the branching fractions of
the \JpsiModes modes, this approach shifts \AI in each bin by
approximately 4\%. The isospin asymmetries are shown in
Fig.~\ref{fig:AI} for \BKMuMu and \BKstMuMu and given in
Tables~\ref{tab:kmumu_AI} and~\ref{tab:kstmumu_AI} in the
appendix. The asymmetric uncertainties are obtained from the profile
likelihood.

Since there is no knowledge on the shape of \AI in models that extend
the SM, apart from large correlations expected between neighbouring
bins, the $\AI=0$ hypothesis is tested against the simplest
alternative, that is a constant value different from zero. The
difference in \chisq between the two hypotheses is used as a test
statistic and is compared to the differences in an ensemble of
pseudo-experiments which are generated with zero isospin
asymmetry. Given the current statistical precision, the hypothesis of
$\AI=0$ is a good approximation to the SM which predicts \AI to be
$\mathcal{O}(1\%)$~\cite{isospintheory,Khodjamirian:2012rm,roman}.
The $p$-value for the \BKMuMu isospin asymmetry under the $\AI=0$
hypothesis is 11\%, corresponding to a significance of
1.5\,$\sigma$. The \BKstMuMu isospin asymmetry has a $p$-value of 80\%
with respect to zero. Alternatively, a simple \chisq test of the data
with respect to a hypothesis of zero isospin asymmetry has a $p$-value
of 54\%~(4\%) for the \BKMuMu~(\BKstMuMu) isospin asymmetry.

\begin{figure}[tb]
\centering
\subfigure{\includegraphics[width=0.45\textwidth] {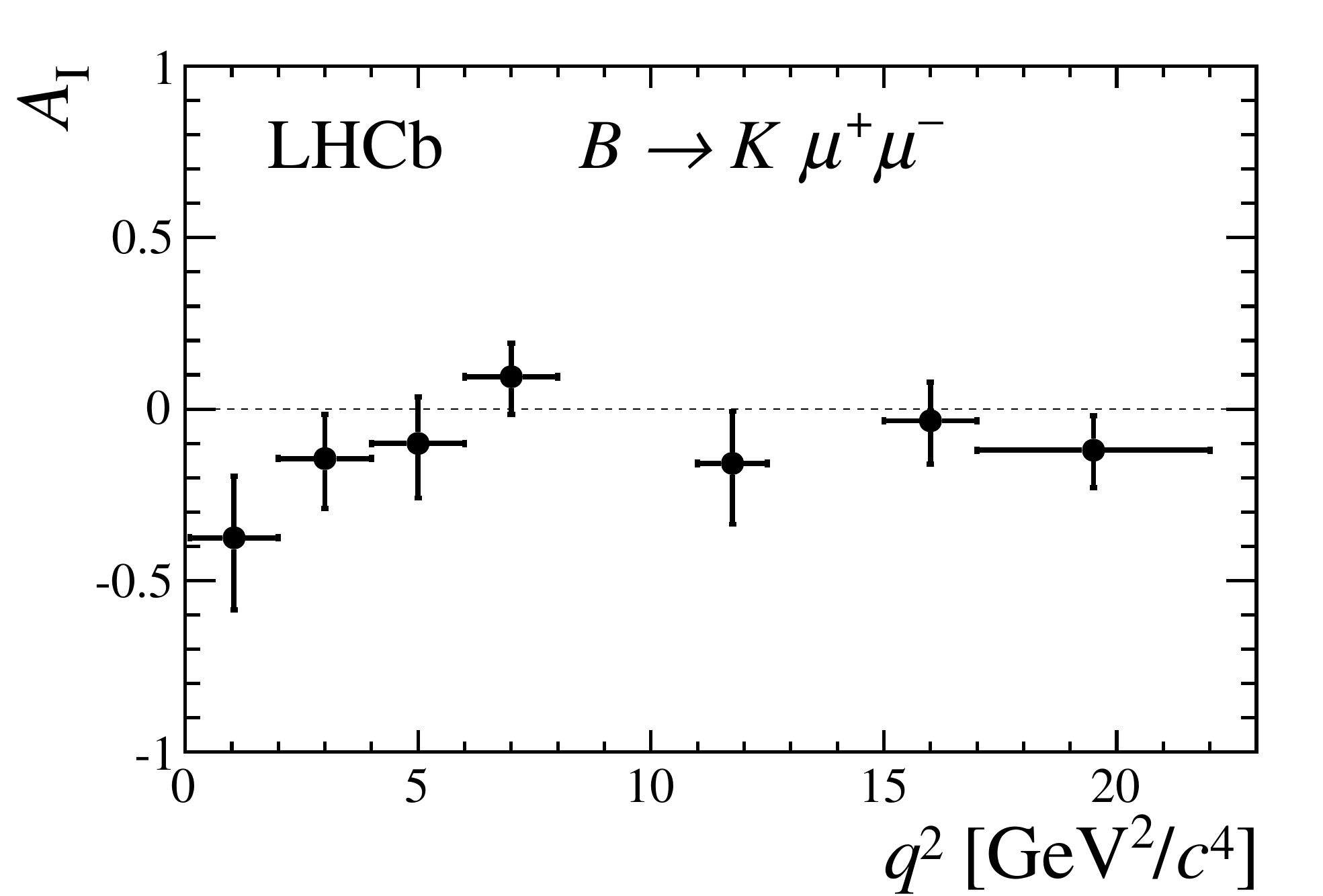}}
\subfigure{\includegraphics[width=0.45\textwidth] {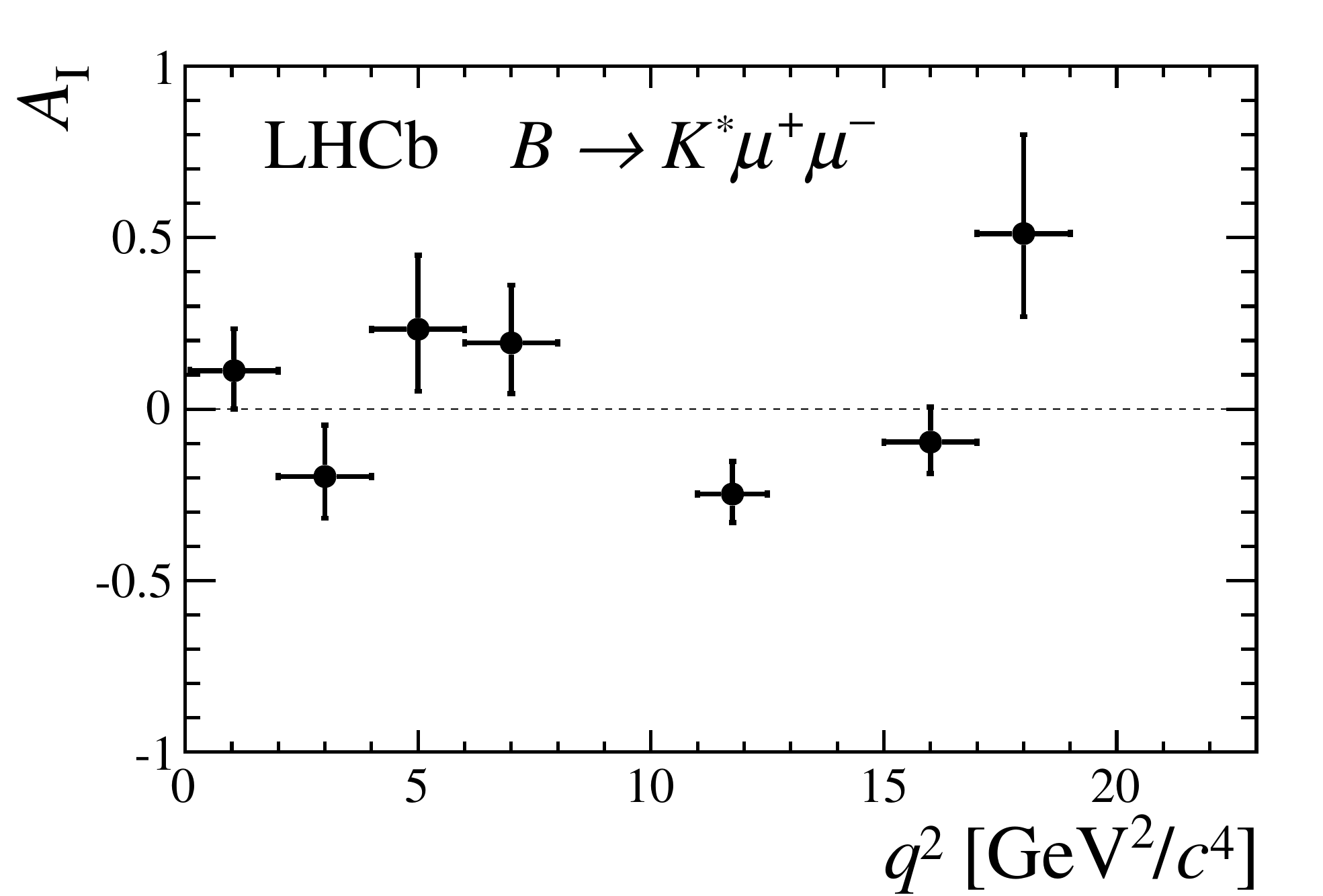}}
\caption{\small Isospin asymmetries for (left) \BKMuMu and (right) \BKstMuMu 
  decays.}
  \label{fig:AI}
\end{figure}

\begin{figure}[tb]
\centering
\includegraphics[width=0.45\textwidth] {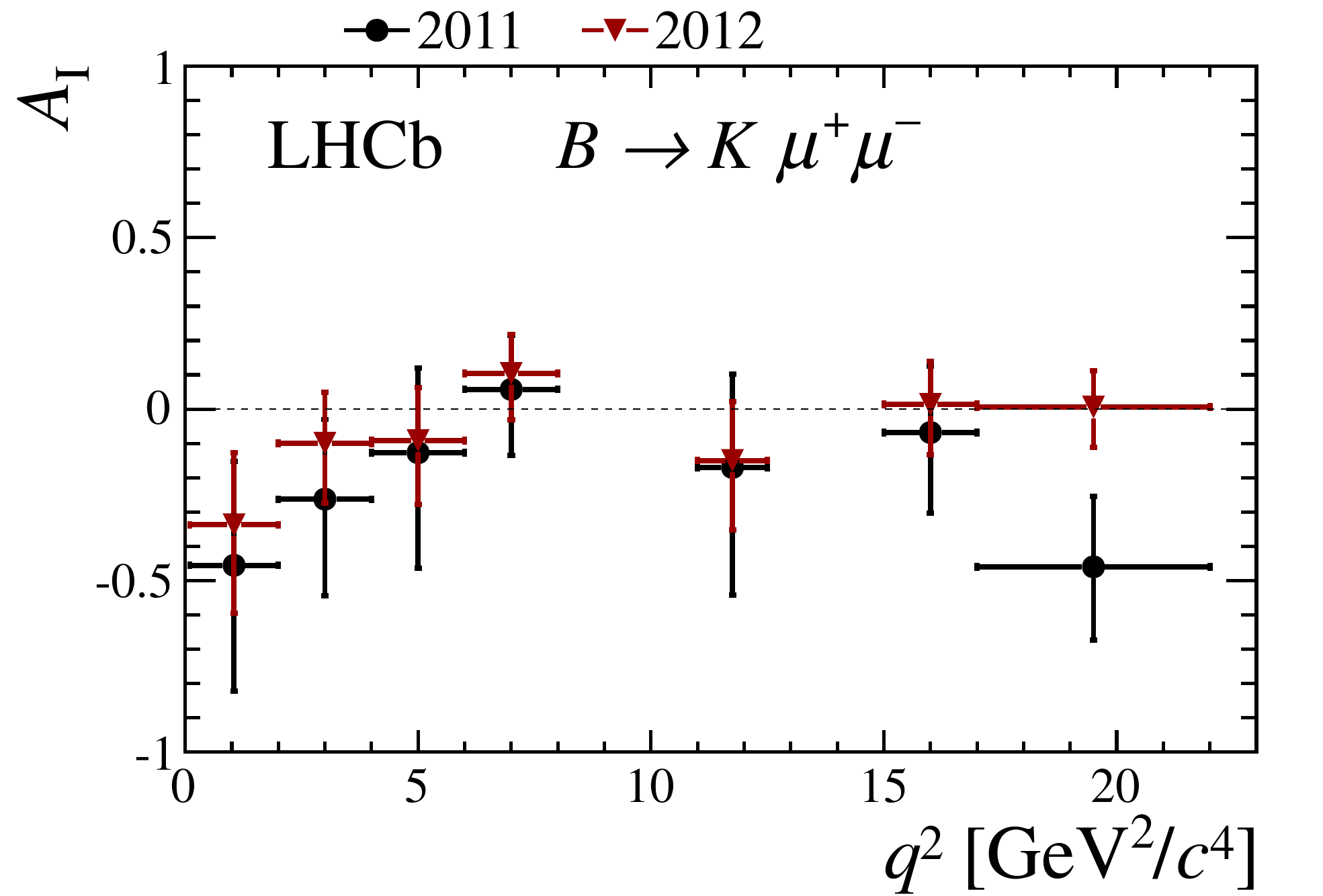}
\caption{\small Isospin asymmetry of \BKMuMu obtained separately from
  the 2011 and 2012 data sets.}
\label{fig:years}
\end{figure}

Although the isospin asymmetry for \BKMuMu decays is negative in all
but one \qq bin, results are more consistent with the SM compared to
the previous measurement in Ref.~\cite{LHCb-PAPER-2012-011}, which
quoted a 4.4\,$\sigma$ significance to differ from zero, using a test
statistic that explicitly tested for \AI to be negative in all
bins. The lower significance quoted here is due to four effects: the
change of the test statistic in the calculation of the significance
itself, which reduces the previous discrepancy to 3.5\,$\sigma$; the
assumption that the isospin asymmetry of \JpsiModes is zero which
reduces the significance further to 3.2\,$\sigma$; a re-analysis of
the 2011 data with the updated reconstruction and event selection that
reduces the significance to 2.5\,$\sigma$; and finally the inclusion
of the 2012 data set reduces the significance further to
1.5\,$\sigma$.

The measurements of \AI in the individual \qq bins obtained from the
re-analysis of the 2011 data set are compatible with those obtained in
the previous analysis; a $\chi^2$ test on the compatibility of the two
results, taking the overlap of events into account, has a $p$-value of
93\%. However results from the 2012 data are more compatible with an
\AI value of zero than the re-analysed 2011 data, as shown in
Fig~\ref{fig:years}.


\section{Conclusion}
\label{sec:Conclusion}

The most precise measurements of the differential branching fractions
of \BuKMuMu, \BdKsMuMu and \BuKstMuMu decays as well as the isospin
asymmetries of \BKMuMu and \BKstMuMu decays have been performed using
a data set corresponding to 3\invfb of integrated luminosity collected
by the LHCb detector.

The isospin asymmetries of the \BKMuMu and \BKstMuMu decays are both
consistent with SM expectations. However, the branching fraction
measurements all have lower values than the SM predictions.  This is consistent with
the $\BdKstMuMu$ and $\Bs\to\phi\mumu$ branching fractions measured by
LHCb, which also favour lower
values~\cite{LHCb-PAPER-2013-019,LHCb-PAPER-2013-017,kstmumulattice2}
than predicted by the SM.

\section*{Acknowledgements} 

\noindent 
We would like to thank Roman Zwicky for the useful discusions
regarding \AI in the \JpsiModes system, and Chris Bouchard and Stefan
Meinel for information on branching fraction predictions of \BKMuMu
and \BKstMuMu from the lattice calculations. We express our gratitude
to our colleagues in the CERN accelerator departments for the
excellent performance of the LHC. We thank the technical and
administrative staff at the LHCb institutes. We acknowledge support
from CERN and from the national agencies: CAPES, CNPq, FAPERJ and
FINEP (Brazil); NSFC (China); CNRS/IN2P3 and Region Auvergne (France);
BMBF, DFG, HGF and MPG (Germany); SFI (Ireland); INFN (Italy); FOM and
NWO (The Netherlands); SCSR (Poland); MEN/IFA (Romania); MinES,
Rosatom, RFBR and NRC ``Kurchatov Institute'' (Russia); MinECo,
XuntaGal and GENCAT (Spain); SNSF and SER (Switzerland); NASU
(Ukraine); STFC and the Royal Society (United Kingdom); NSF (USA). We
also acknowledge the support received from EPLANET, Marie Curie
Actions and the ERC under FP7.  The Tier1 computing centres are
supported by IN2P3 (France), KIT and BMBF (Germany), INFN (Italy), NWO
and SURF (The Netherlands), PIC (Spain), GridPP (United Kingdom).  We
are indebted to the communities behind the multiple open source
software packages on which we depend.  We are also thankful for the
computing resources and the access to software R\&D tools provided by
Yandex LLC (Russia).

\addcontentsline{toc}{section}{References}
\ifx\mcitethebibliography\mciteundefinedmacro
\PackageError{LHCb.bst}{mciteplus.sty has not been loaded}
{This bibstyle requires the use of the mciteplus package.}\fi
\providecommand{\href}[2]{#2}

\newpage
\begin{appendix}
\section{Appendix}
\label{sec:Appendix}

\begin{table}[ht]
  \centering
  \caption{Differential branching fraction results ($10^{-9} \times c^4/\mbox{GeV}^2$) for the
    \BuKMuMu decay, including statistical and systematic uncertainties.}
    \begin{tabular}{c c c c}
      \hline
      \noalign{\vskip 1mm}
      \qq range (\!$\gev^{2}/c^{4}$)& central value & stat & syst \\
      \hline
      \noalign{\vskip 1mm}
0.1 $<$ \qq $<$ 0.98 & 33.2 & 1.8 & 1.7 \\
1.1 $<$ \qq $<$ 2.0 & 23.3 & 1.5 & 1.2 \\
2.0 $<$ \qq $<$ 3.0 & 28.2 & 1.6 & 1.4 \\
3.0 $<$ \qq $<$ 4.0 & 25.4 & 1.5 & 1.3 \\
4.0 $<$ \qq $<$ 5.0 & 22.1 & 1.4 & 1.1 \\
5.0 $<$ \qq $<$ 6.0 & 23.1 & 1.4 & 1.2 \\
6.0 $<$ \qq $<$ 7.0 & 24.5 & 1.4 & 1.2 \\
7.0 $<$ \qq $<$ 8.0 & 23.1 & 1.4 & 1.2 \\
11.0 $<$ \qq $<$ 11.8 & 17.7 & 1.3 & 0.9 \\
11.8 $<$ \qq $<$ 12.5 & 19.3 & 1.2 & 1.0 \\
15.0 $<$ \qq $<$ 16.0 & 16.1 & 1.0 & 0.8 \\
16.0 $<$ \qq $<$ 17.0 & 16.4 & 1.0 & 0.8 \\
17.0 $<$ \qq $<$ 18.0 & 20.6 & 1.1 & 1.0 \\
18.0 $<$ \qq $<$ 19.0 & 13.7 & 1.0 & 0.7 \\
19.0 $<$ \qq $<$ 20.0 & \phantom{0}7.4 & 0.8 & 0.4 \\
20.0 $<$ \qq $<$ 21.0 & \phantom{0}5.9 & 0.7 & 0.3 \\
21.0 $<$ \qq $<$ 22.0 & \phantom{0}4.3 & 0.7 & 0.2 \\
\hline
      \noalign{\vskip 1mm}
1.1 $<$ \qq $<$ 6.0 & 24.2 & 0.7 & 1.2 \\
15.0 $<$ \qq $<$ 22.0 & 12.1 & 0.4 & 0.6 \\
      \hline
    \end{tabular}
  \label{tab:kmumu_results}
  \end{table}
  
  \begin{table}[ht]
  \centering
   \caption{Differential branching fraction results ($10^{-9} \times c^4/\mbox{GeV}^2$) for the \BdKzMuMu decay, including statistical and systematic uncertainties.}
  \setlength{\extrarowheight}{3pt}
    \begin{tabular}{c c c c}
      \hline
      \qq range (\!$\gev^{2}/c^{4}$) & central value & stat & syst \\
      \hline
0.1 $<$ \qq $<$ 2.0 & 12.2 & $^{+5.9}_{-5.2}$ & 0.6 \\
2.0 $<$ \qq $<$ 4.0 & 18.7 & $^{+5.5}_{-4.9}$ & 0.9 \\
4.0 $<$ \qq $<$ 6.0 & 17.3 & $^{+5.3}_{-4.8}$ & 0.9 \\
6.0 $<$ \qq $<$ 8.0 & 27.0 & $^{+5.8}_{-5.3}$ & 1.4 \\
11.0 $<$ \qq $<$ 12.5 & 12.7 & $^{+4.5}_{-4.0}$ & 0.6 \\
15.0 $<$ \qq $<$ 17.0 & 14.3 & $^{+3.5}_{-3.2}$ & 0.7 \\
17.0 $<$ \qq $<$ 22.0 & \phantom{0}7.8 & $^{+1.7}_{-1.5}$ & 0.4 \\
\hline
1.1 $<$ \qq $<$ 6.0 & 18.7 & $^{+3.5}_{-3.2}$ & 0.9\\
15.0 $<$ \qq $<$ 22.0 & \phantom{0}9.5 & $^{+1.6}_{-1.5}$ & 0.5\\
      \hline
    \end{tabular}
  \label{tab:ksmumu_results}
  \end{table}

\begin{table}[ht]
  \centering
   \caption{Differential branching fraction results ($10^{-9} \times c^4/\mbox{GeV}^2$) for the \BuKstMuMu decay, including statistical and systematic uncertainties.}
 \setlength{\extrarowheight}{3pt}
    \begin{tabular}{c c c c}
      \hline
      \qq range (\!$\gev^{2}/c^{4}$) & central value & stat & syst \\
      \hline
0.1 $<$ \qq $<$ 2.0 & 59.2 & $^{+14.4}_{-13.0}$ & 4.0 \\
2.0 $<$ \qq $<$ 4.0 & 55.9 & $^{+15.9}_{-14.4}$ & 3.8 \\
4.0 $<$ \qq $<$ 6.0 & 24.9 & $^{+11.0}_{-\phantom{0}9.6}$ & 1.7 \\
6.0 $<$ \qq $<$ 8.0 & 33.0 & $^{+11.3}_{-\phantom{0}10.0}$ & 2.3 \\
11.0 $<$ \qq $<$ 12.5 & 82.8 & $^{+15.8}_{-14.1}$ & 5.6 \\
15.0 $<$ \qq $<$ 17.0 & 64.4 & $^{+12.9}_{-11.5}$ & 4.4 \\
17.0 $<$ \qq $<$ 19.0 & 11.6 & $^{+\phantom{0}9.1}_{-\phantom{0}7.6}$ & 0.8 \\
\hline
1.1 $<$ \qq $<$ 6.0 & 36.6 & $^{+\phantom{0}8.3}_{-\phantom{0}7.6}$ & 2.6\\
15 $<$ \qq $<$ 19.0 & 39.5 & $^{+\phantom{0}8.0}_{-\phantom{0}7.3}$ & 2.8\\
      \hline
    \end{tabular}
  \label{tab:bukst_results}
  \end{table}

 \begin{table}[ht]
  \centering
   \caption{Isospin asymmetry results for the \BKMuMu decay, including statistical and systematic uncertainties.}
 \setlength{\extrarowheight}{3pt}
    \begin{tabular}{c c c c}
      \hline
      \qq range (\!$\gev^{2}/c^{4}$) & central value & stat & syst \\
      \hline
0.1 $<$ \qq $<$ 2.0 & -0.37 & $^{+0.18}_{-0.21}$ & 0.02\\
2.0 $<$ \qq $<$ 4.0 & -0.15 & $^{+0.13}_{-0.15}$ & 0.02\\
4.0 $<$ \qq $<$ 6.0 & -0.10 & $^{+0.13}_{-0.16}$ & 0.02\\
6.0 $<$ \qq $<$ 8.0 & \phantom{-}0.09 & $^{+0.10}_{-0.11}$ & 0.02\\
11.0 $<$ \qq $<$ 12.5 & -0.16 & $^{+0.15}_{-0.18}$ & 0.03\\
15.0 $<$ \qq $<$ 17.0 & -0.04 & $^{+0.11}_{-0.13}$ & 0.02\\
17.0 $<$ \qq $<$ 22.0 & -0.12 & $^{+0.10}_{-0.11}$ & 0.02\\
\hline
1.1 $<$ \qq $<$ 6.0 & -0.10 & $^{+0.08}_{-0.09}$ & 0.02\\
15.0 $<$ \qq $<$ 22.0 & -0.09 & $^{+0.08}_{-0.08}$ & 0.02\\
      \hline
    \end{tabular}
  \label{tab:kmumu_AI}
  \end{table}

\begin{table}[ht]
  \centering
   \caption{Isospin asymmetry results for the \BKstMuMu decay, including statistical and systematic uncertainties.}
 \setlength{\extrarowheight}{3pt}
    \begin{tabular}{c c c c}
      \hline
      \qq range (\!$\gev^{2}/c^{4}$) & central value & stat & syst \\
      \hline
0.1 $<$ \qq $<$ 2.0 & \phantom{-}0.11 & $^{+0.12}_{-0.11}$ & 0.02\\
2.0 $<$ \qq $<$ 4.0 & -0.20 & $^{+0.15}_{-0.12}$ & 0.03\\
4.0 $<$ \qq $<$ 6.0 & \phantom{-}0.23 & $^{+0.21}_{-0.18}$ & 0.02\\
6.0 $<$ \qq $<$ 8.0 & \phantom{-}0.19 & $^{+0.17}_{-0.15}$ & 0.02\\
11.0 $<$ \qq $<$ 12.5 & -0.25 & $^{+0.09}_{-0.08}$ & 0.03\\
15.0 $<$ \qq $<$ 17.0 & -0.10 & $^{+0.10}_{-0.09}$ & 0.03\\
17.0 $<$ \qq $<$ 19.0 & \phantom{-}0.51 & $^{+0.29}_{-0.24}$ & 0.02\\
\hline
1.1 $<$ \qq $<$ 6.0 & \phantom{-}0.00 & $^{+0.12}_{-0.10}$ & 0.02\\
15.0 $<$ \qq $<$ 19.0 & \phantom{-}0.06 & $^{+0.10}_{-0.09}$ & 0.02\\
      \hline
    \end{tabular}
  \label{tab:kstmumu_AI}
  \end{table}
  
\end{appendix}

\newpage

\end{document}